\documentstyle[prl,aps,psfig,multicol]{revtex}
\begin{document}
\draft

\title{Nonlocality in mesoscopic Josephson junctions with strip geometry}
\author{Urs Ledermann, Alban L. Fauch\`{e}re, and Gianni Blatter}
\address{Theoretische Physik, Eidgen\"{o}ssische Technische Hochschule,
CH-8093 Z\"{u}rich, Switzerland}

\date{\today}
\maketitle

\begin{abstract}
{We study the current in a clean superconductor--normal-metal--superconductor
junction of length $d$ and width $w$ in the presence of an applied magnetic
field $H$. We show that both the geometrical pattern of the current density
and the critical current $I_{c}(\Phi)$, where $\Phi$ is the total flux in the
junction, depend on the ratio of the Josephson vortex distance
$a_{\scriptscriptstyle{0}}=\Phi_{0}/Hd$ and the range
$r\sim\sqrt{d\xi_{\scriptscriptstyle{N}}}$ of the nonlocal electrodynamics
($\Phi_{0}=hc/2e$, $\xi_{\scriptscriptstyle{N}}=\hbar
v_{\scriptscriptstyle{F}}/2\pi T$, and $r(T\rightarrow 0)\sim d$). In
particular, the critical current has the periodicity of the superconducting
flux quantum $\Phi_{0}$ only for $r<a_{\scriptscriptstyle{0}}$ and becomes,
due to boundary effects, $2\Phi_{0}$ (pseudo-) periodic for strong
nonlocality, $r>a_{\scriptscriptstyle{0}}$. Comparing our results to recent
experiments of Heida \textit{et al.} [Phys. Rev. B \textbf{57}, R5618 (1998)]
we find good agreement.}
\end{abstract}


\begin{multicols}{2}  

\narrowtext

Quantum mechanical interference effects render the electrodynamics of
mesoscopic samples \emph{nonlocal}. In particular, nonlocality is a key
element entering the understanding of the magnetic response and the transport
in SNS-junctions ($s$-wave superconductor--normal-metal--$s$-wave
superconductor junctions) and SN-proximity sandwiches. The strength and
relevance of the nonlocality in general depends on the dimensions of the
system, the normal-metal coherence length $\xi_{\scriptscriptstyle{N}}=\hbar
v_{\scriptscriptstyle{F}}/2\pi T$, and the elastic scattering length $l$
\cite{bib:Kulik,bib:Zaikin,bib:Belzig}. In this paper we show, how the
different length scales enter the magnetotransport problem of a mesoscopic
SNS-junction to produce a shift in the (pseudo-) periodicity of the critical
current from $\Phi_{0}$ to~$2\Phi_{0}$.

\begin{figure}[h]
  \centerline{
  \psfig{file=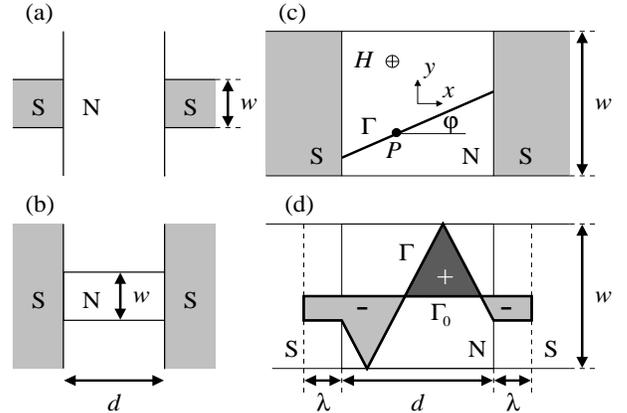,width=8cm,height=5.5cm}}
  \vspace{4mm} \caption{(a) Junction with a point-contact (open) geometry
  as discussed in [10], where $w$ is the width of the two
  \emph{super}conductors. (b) The junction studied here has a strip geometry
  with $w$ the width of the \emph{normal} conductor. (c) The magnetic field
  $H$ is applied in the $z$-direction and the coordinate system is chosen
  symmetric with respect to the junction center. The current density in the
  point $P$ involves contributions from all trajectories $\Gamma$
  parameterized by the angle $\varphi$. (d) The phase difference $\gamma$
  along the trajectory $\Gamma$ can be expressed through the enclosed flux
  $\phi$; using the trajectory $\Gamma_{0}$ as our reference, the flux through
  the areas above (below) $\Gamma_{0}$ contributes with a positive (negative)
  sign.}  \label{f:SNSJunction}
\end{figure}

After the discovery of the Josephson effect in SIS tunnel
junctions~\cite{bib:Josephson}, interest turned to metallic links of the SNS
type~\cite{bib:Likharev}, where the current is conveniently described in terms
of Andreev states trapped within the normal-metal region
\cite{bib:Andreev}. For a wide junction, the current density and the
supercurrent in the presence of a magnetic field $H$ have been calculated by
Antsygina \textit{et al.}~\cite{bib:Svidz}, who found a $\Phi_{0}$
periodicity in the critical current. Continuous progress in nanofabrication
technology made it possible to investigate mesoscopic
superconductor--semiconductor heterostructures, see Ref. \cite{bib:Taka} for a
study of the fluctuations in the critical current and its quantization in a
superconducting quantum point contact. Recently, Heida \textit{et al.}
\cite{bib:Heida}, investigating S-2DEG-S-junctions ($s$-wave
superconductor--two-dimensional-electron-gas--$s$-wave superconductor
junctions) of comparable width $w$ and length $d$, have measured a $2\Phi_{0}$
periodicity of the critical current instead of the standard $\Phi_{0}$. A
first attempt to explain this finding is due to Barzykin and Zagoskin
\cite{bib:Zagoskin}; considering the point-contact geometry of Fig.~1(a) with
open boundary condition in the metal, they indeed recover a $2\Phi_{0}$
periodicity in the limit $w/d\rightarrow 0$. However, the experiment of Heida
\textit{et al.} \cite{bib:Heida} is carried out in the strip geometry of
Fig.~1(b) and involves dimensions $w\sim d$ of the same order. In the present
paper, we determine the critical current $I_{c}$ through a clean SNS-junction
in the presence of an applied magnetic field $H$, taking proper account of the
reflecting boundaries in the normal-metal characteristic for the strip
geometry of Fig.~\ref{f:SNSJunction}(b).

We find that the periodicity of the critical current changes from $\Phi_{0}$
to $2\Phi_{0}$ as the flux through the junction increases. At low temperatures
the crossover to the $2\Phi_{0}$ periodic current appears at a flux
$\sim\Phi_{0} w/d$, thus explaining the result of Heida {\it et al.}
\cite{bib:Heida}, who found a $2\Phi_{0}$ periodic pattern for all fields in
devices with $w/d\sim 1$.

In our derivation, we neglect screening effects by the induced supercurrent,
which is justified as long as $H>\sqrt{8\Phi_{0} j_{c}/cd^{\prime}}$, where
$j_{c}$ denotes the critical current density of the junction and
$d^{\prime}=d+2\lambda$ with $\lambda$ the penetration depth of the two bulk
superconductors~\cite{bib:Tinkham}. Since all the length scales in our system
(the dimensions $w$ and $d$, the normal-metal coherence length
$\xi_{\scriptscriptstyle{N}}$) are much larger than the Fermi wave length
$\lambda_{F}$, we base our calculations on the Eilenberger equations
\cite{bib:Eilenberger} for the quasi-classical Green functions and extract the
current density in the standard way~\cite{bib:Belzig}.

The SNS-junction we study is sketched in Fig.~\ref{f:SNSJunction}. In the
quasi-classical formulation, the current density in a point $P$ results from
contributions over all \emph{trajectories} of quasiparticles connecting one
interface to the other through $P$. In a junction of infinite width, all
trajectories go straight through the junction. In the case of a finite
junction, boundary conditions at the normal-metal--vacuum boundary have to be
applied, which we idealize through the assumption of specular
reflections. Furthermore, we adopt the usual approximations: perfect Andreev
reflections at the SN-interfaces and a coherence length $\xi_{0}$ in the two
superconductors with $\xi_{0}\ll d$, allowing for a step-like approximation of
the order parameter~$\Delta$ \cite{bib:Kulik,bib:Svidz,bib:Ishii}. The
quasi-classical Green function is calculated by matching the partial solutions
in N and S at the interfaces. For the current density $\mathbf{j}$, we arrive
at a generalization of the results of Antsygina \textit{et
al.}~\cite{bib:Svidz}. Explicitely, for finite temperatures with
$d\gg\xi_{\scriptscriptstyle{N}}$, $\mathbf{j}$ takes the form,
\begin{eqnarray}
  \frac{{\bf j}(x,y)}{j_{c,T}}=\frac{-1}{\sqrt{2\pi}}
  \int_{-\pi/2}^{\pi/2}\!\!\!d\varphi\,\hat{\mathbf{p}}\,
  \frac{\sin(\gamma)d}{\sqrt{\xi_{\scriptscriptstyle{N}} l(\varphi)}}
  \exp\left(\frac{d-l(\varphi)}{\xi_{\scriptscriptstyle{N}}}\right),
  \label{eq:CurrentDensHighTemp}
\end{eqnarray}
while in the low temperature limit, $d\ll\xi_{\scriptscriptstyle{N}}$,
\begin{eqnarray}
  \frac{{\bf j}(x,y)}{j_{c,0}}=\frac{4}{\pi^{2}}\sum_{k=1}^{\infty}
  \frac{(-1)^{k}}{k}\!\int_{-\pi/2}^{\pi/2}\!\!\!
  d\varphi\;\hat{\mathbf{p}}\sin(k\gamma)\frac{d}{l(\varphi)},
  \label{eq:CurrentDensLowTempST}
\end{eqnarray}
where $\hat{\mathbf{p}}=(\cos(\varphi),\sin(\varphi),0)$,
$l(\varphi)=d/\cos(\varphi)$ is the length of a trajectory with slope
$\varphi$, and the critical current densities are
\begin{equation}
  j_{c,T}=\rho
  j_{c,0}\exp\left(-\frac{d}{\xi_{\scriptscriptstyle{N}}}\right)\;\;{\rm
  and}\;\; j_{c,0}=\frac{ne^{2}}{mc}\frac{\Phi_{0}}{2d}.  \label{eq:cd}
\end{equation}
In (\ref{eq:cd}), $n$ denotes the electron density in the normal conductor and
$\rho\approx 12/\pi$ for $T\ll T_{c}$, $\lim_{T\rightarrow T_{c}}\rho\sim
1-T/T_{c}$. While in the low temperature limit all harmonics $\sin(k\gamma)$
($k=1,2,\ldots$) contribute to the current density \cite{bib:Ishii}, at finite
temperatures, thermal smearing of the Andreev levels leads to a suppression of
the higher harmonics $\propto\exp(-kd/\xi_{\scriptscriptstyle{N}})$ and only
the first term $\propto\sin(\gamma)$ survives. An individual trajectory
contributes with a weight $\propto\exp(-l/\xi_{\scriptscriptstyle{N}})$ at
finite- and $\propto d/l$ in the low temperature limit. In a wide junction,
$\gamma$ takes the form~\cite{bib:Svidz},
\begin{equation}
  \gamma(x,y;\varphi)=\gamma_{0}-\frac{2\pi}{\Phi_{0}}Hd^{\prime}
  \left[y-x\tan(\varphi)\right].  \label{eq:GammaInf}
\end{equation}
The more general expression derived here results in the gauge invariant
phase difference
\begin{equation}
  \gamma(x,y;\varphi)=\Delta\varphi-\frac{2\pi}{\Phi_{0}} \int_{\Gamma}{\bf
  A}\cdot d{\bf s}, \label{eq:GaugeInvPhd}
\end{equation}
where $\Delta\varphi$ denotes the phase difference between the two
superconductors and $\Gamma$ is the path which goes through the point $(x,y)$
with slope $\varphi$. Combining the current expressions
(\ref{eq:CurrentDensHighTemp}) or (\ref{eq:CurrentDensLowTempST}) and
(\ref{eq:GaugeInvPhd}) with the Maxwell equation $\nabla^{2}{\bf A}=-4\pi{\bf
j}[{\bf A},\Delta\varphi]/c$, we obtain the transverse vector-potential ${\bf
A}$ and thus can solve the full screening problem; in the case of a tunnel
junction (where the trajectories are reduced to the one with $\varphi=0$),
numeric and analytic calculations have been given by Owen and
Scalapino~\cite{bib:Scalapino}.

In the following, we neglect screening and concentrate on junctions with the
strip geometry of Fig.~\ref{f:SNSJunction}(b), including the (reflecting)
trajectories $\Gamma$ in (\ref{eq:CurrentDensHighTemp}) and
(\ref{eq:CurrentDensLowTempST}). We express the gauge invariant phase
difference (\ref{eq:GaugeInvPhd}) in terms of the flux $\phi$ enclosed by
$\Gamma$ and the reference path $\Gamma_{0}$ and obtain,
\begin{equation}
  \gamma(x,y;\varphi)=\gamma_{0}-\frac{2\pi\phi(x,y;\varphi)}{\Phi_{0}},
\end{equation}
where for negligible screening $\phi(x,y;\varphi)=H{\mathcal A}(x,y;\varphi)$
and $\mathcal{A}$ is the properly weighted enclosed area, see
Fig.~\ref{f:SNSJunction}(d). The area $\mathcal{A}$ is calculated as a
function of the number of reflections the trajectory $\Gamma$ undergoes (in
the following called the `order' of the trajectory). The point-contact
geometry of Fig.~\ref{f:SNSJunction}(a) then is described by the order-zero
trajectories alone \cite{bib:Zagoskin}, while in the strip geometry of
Fig.~\ref{f:SNSJunction}(b), higher orders have to be included.

\begin{figure}[ht]
  \centerline{\psfig{file=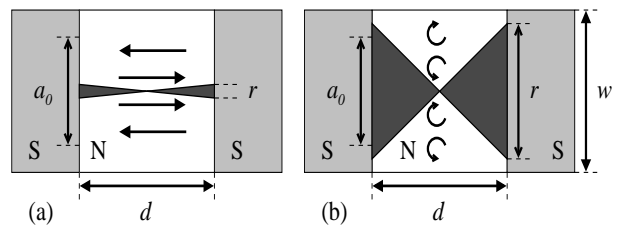,width=8cm,height=2.9cm}}
  \vspace{4mm} \caption{The `bow ties' of width $r$ display the ensemble of
  trajectories contributing to the current density in a given point. The
  arrows indicate the current flow. (a) For weak nonlocality,
  $r<a_{\scriptscriptstyle{0}}$, the current flows straight through the
  junction. (b) The strong nonlocality, $r>a_{\scriptscriptstyle{0}}$, leads
  to the formation of vortex-like domains of circular flow.}
  \label{f:nlrange}
\end{figure}
\begin{figure}[ht]
    \psfig{file=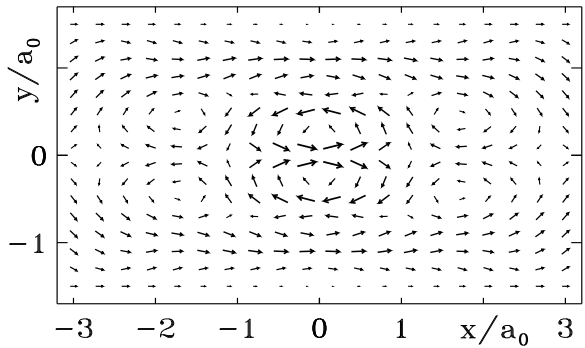,width=8.5cm,height=5cm}
    \caption{The current density for $d/\xi_{\scriptscriptstyle{N}}=3$,
    $r/a_{\scriptscriptstyle{0}}\approx 3.5$, and $\Phi/\Phi_{0}=3$. Apart
    from the vortex-like domains in the middle produced by order-zero
    trajectories, additional circular flow is set up by the order-one
    trajectories.}  \label{f:cdnl}
\end{figure}

The geometrical pattern in the current density $\mathbf{j}$ depends strongly
on the sample dimensions $d$ and $w$, the normal-metal coherence length
$\xi_{\scriptscriptstyle{N}}$, and the applied field $H$. At finite
temperature, the current density in $P$ draws its weight from trajectories
with $\varphi<\sqrt{\xi_{\scriptscriptstyle{N}}/d}$, allowing us to introduce
the transverse nonlocality range $r=\sqrt{\xi_{\scriptscriptstyle{N}} d}$ (in
the low temperature limit, $\varphi\sim 1$ and we define $r=d$). This range of
nonlocality has to be compared to the scale
$a_{\scriptscriptstyle{0}}=\Phi_{0}/Hd^{\prime}$ of transverse variations in
$\mathbf{j}$ (see Fig.~\ref{f:nlrange}): For \emph{weak} nonlocality,
$r<a_{\scriptscriptstyle{0}}$, the flow is uniform along $x$ with amplitude
$j_{c}$ and changes direction on a distance $a_{\scriptscriptstyle{0}}/2$
going up the $y$-axis. This contrasts with the \emph{strongly} nonlocal case
$r>a_{\scriptscriptstyle{0}}$, where the current density forms domains of left
and right going circular flow. While the local case is similar to that in a
tunnel junction, the pattern in the \emph{non}local situation reminds of the
usual vortex structure in a superconductor, see
Fig. \ref{f:cdnl}. Explicitely, for finite temperatures with
$d\gg\xi_{\scriptscriptstyle{N}}$, the current density of the order-zero
trajectories is given by
\begin{eqnarray}
  \frac{j_{x}(x,y)}{j_{c,T}}&=&-\sin\left(\gamma_{0}-\frac{2\pi
  y}{a_{\scriptscriptstyle{0}}}\right) \exp\left[-\alpha(x)\right],
  \nonumber\\ \frac{j_{y}(x,y)}{j_{c,T}}&=&-\cos\left(\gamma_{0}-\frac{2\pi
  y}{a_{\scriptscriptstyle{0}}}\right)
  \frac{\xi_{\scriptscriptstyle{N}}}{d}\frac{2\pi
  x}{a_{\scriptscriptstyle{0}}}\exp\left[-\alpha(x)\right],
\end{eqnarray}
where
\begin{equation}
  \alpha(x)=-\frac{d}{\xi_{\scriptscriptstyle{N}}}+
  \sqrt{\left(\frac{d}{\xi_{\scriptscriptstyle{N}}}\right)^{2}
  +\left(\frac{2\pi x}{a_{\scriptscriptstyle{0}}}\right)^{2}}.
\end{equation}
For weak nonlocality, $a_{\scriptscriptstyle{0}}<r$, the exponent remains
approximately constant in the normal part, $\alpha(x)\approx\alpha(0)$,
leading to a uniform current flow, while for strong nonlocality,
$a_{\scriptscriptstyle{0}}>r$, $\alpha(x)$ grows as $x$ approaches the
interfaces, $\alpha(\pm d/2)\gg\alpha(0)$, such that the current concentrates
in the middle of the junction. For $a_{\scriptscriptstyle{0}}>r$, the higher
order trajectories lead to a refinement of the current pattern, see
Fig. \ref{f:cdnl} (similar results are obtained in the low temperature limit).

\begin{figure}[h]
  \psfig{file=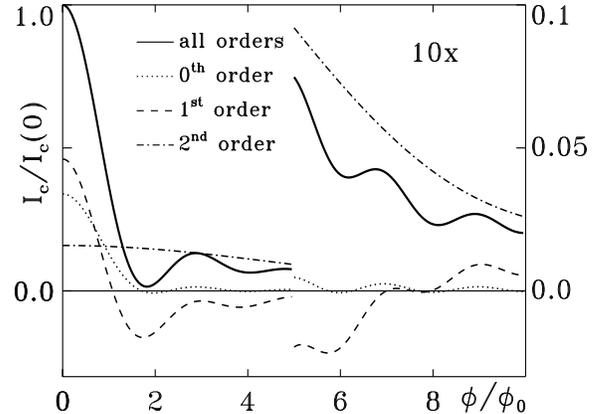,width=7.5cm,height=6cm}
  \caption{The critical current for $d/\xi_{\scriptscriptstyle{N}}=5$ and
  $w/d=1/3$. The solid curve shows the full critical current and the dashed
  curves are the contributions from the orders 0, 1, and 2. The orders 0 and 1
  oscillate with periodicity $2\Phi_{0}$, while the second order decreases
  monotonically, remaining always positive. The current pattern produced by
  the orders 0 and 1 is lifted by the order 2 contributions, and the critical
  current attains the periodicity $2\Phi_{0}$.}  \label{f:CritCurrentLift}
\end{figure}

The ratio $r/a_{\scriptscriptstyle{0}}$ and its associated characteristic
current pattern manifest themselves in the (pseudo-) periodicity of the
critical current,
\begin{equation}
  I_{c}(\Phi)=\max_{\gamma_{0}}\int_{-w/2}^{w/2}dy\;
  j_{x}(0,y;\gamma_{0},\Phi),
\end{equation}
versus flux $\Phi=Hd^{\prime}w$ in the junction. In the case of weak
nonlocality, $r<a_{\scriptscriptstyle{0}}$, the relevant contribution to the
critical current comes from the order-zero trajectories resulting in a
$\Phi_{0}$ periodicity. For strong nonlocality, $r>a_{\scriptscriptstyle{0}}$,
higher orders are relevant and \emph{lift} the order-zero result as shown in
Fig.~\ref{f:CritCurrentLift} --- the periodicity of the critical current
changes to $2\Phi_{0}$. To be specific, we discuss in detail the orders 0, 1,
and 2 for the case of finite temperatures with
$d\gg\xi_{\scriptscriptstyle{N}}$ (the qualitative arguments for
$d\ll\xi_{\scriptscriptstyle{N}}$ are similar).

For $w>r>a_{\scriptscriptstyle{0}}$, the critical current due to the
order-zero trajectories takes the form~\cite{bib:rb},
\begin{equation}
  I_{c}^{(0)}(\Phi)=-\frac{\sqrt{2}I_{c,T}}{\sqrt{\pi}}\frac{w}{r}
  \frac{\cos(\pi\Phi/\Phi_{0})}{(\pi\Phi/\Phi_{0})^{2}},
\end{equation}
where $I_{c,T}=wj_{c,T}$. For the first-order trajectories, we numerically
find again a $2\Phi_{0}$ (pseudo-) periodic contribution. Both components
vanish with field $\propto 1/\Phi^2$. The second- and all following even-order
trajectories exhibit a large current amplitude of order $j_{c}$ on a scale
$a_{\scriptscriptstyle{0}}\propto 1/\Phi$ in the junction center $(0,0)$, a
consequence of the $\varphi$-independence of the gauge invariant phase
difference $\gamma$ along trajectories through $(0,0)$. Their contribution
scales $\propto 1/\Phi$ and therefore dominates over the zeroth- and
first-order terms at large enough fields --- as the strongly nonlocal limit
with $a_{\scriptscriptstyle{0}}<r$ is reached, the periodicity changes over to
$2\Phi_{0}$.  Samples with a small width $w<r$ are always in the strongly
nonlocal limit and their current pattern is $2\Phi_{0}$ periodic throughout
the entire field axis. At low temperatures, the condition $w<r$ transforms
into the geometric requirement $w < d$. In the very limit $w/d\rightarrow 0$
the periodic modulation disappears and the solution goes over into the zero
field result, $I_{c}(\Phi)\rightarrow I_{c}(0)$. The complete classification
is given in Table~I.

\begin{figure}[h]
  \psfig{file=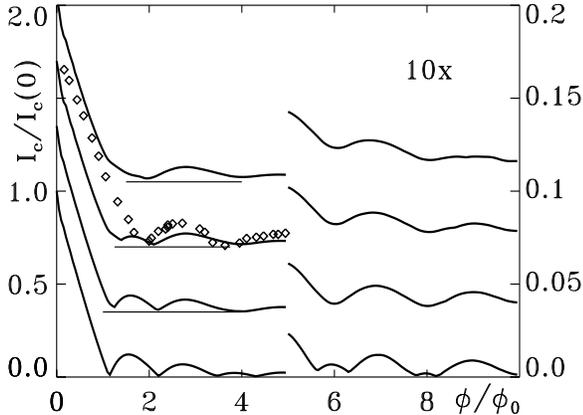,width=7.5cm,height=6cm}
  \caption{The critical current in the low temperature limit,
  $d\ll\xi_{\scriptscriptstyle{N}}$, for $w/d=0.8$, 0.9, 1.1, 1.5 (from top to
  bottom). Successive plots have been offset by 0.35. Measured data ([9],
  diamonds) is shown for the case $w/d=0.9$.}
  \label{f:MaxScLT}
\end{figure}
Recently, Heida {\it et al.} \cite{bib:Heida} observed such a $2\Phi_{0}$
periodicity in strip like ($w\sim d$) S-2DEG-S junctions made from Nb
electrodes in contact with InAs operating at low temperatures $T = 0.1$ K
(similar junctions have been constructed by Takayanagi {\it et al.}, see
Ref. \cite{bib:Taka}). As the total flux through the junction is difficult to
determine in the experiment, Heida {\it et al.} had to infer their $2\Phi_{0}$
periodic structure from a fit on four samples with different ratios $w/d$
ranging from 0.9 to 2.2. In Fig.~\ref{f:MaxScLT}, we present the results of
our numeric calculations for the strip geometry, where we have properly taken
into account the finite penetration depth of the flux into the superconducting
banks. While geometries with $w/d<1$ clearly exhibit a $2\Phi_{0}$ periodicity
throughout the entire field region, a $\Phi_{0}$-component starts to develop
at low fields in wide junctions. The comparison with the data of Heida {\it et
al.} ($w/d = 0.9$) gives a satisfactory description of the pseudo-periodic
structure.

In conclusion, we have demonstrated that the current density and the critical
current in a clean SNS-junction with strip geometry depend crucially on the
ratio $r/a_{\scriptscriptstyle{0}}$ between the nonlocality range $r$ and the
vortex distance $a_{\scriptscriptstyle{0}}$. The period of the critical
current depends not only on the dimensions of the junction and the normal-metal
coherence length $\xi_{\scriptscriptstyle{N}}$, as it is the case in a
point-contact geometry \cite{bib:Zagoskin}, but also on the applied magnetic
field $H$. In particular, we obtain a $2\Phi_{0}$ periodicity in the whole
current pattern at $w\sim d$, whereas for a point contact geometry, the
$2\Phi_{0}$ periodicity is reached only in the limit $w/d\rightarrow 0$
\cite{bib:Zagoskin}. The numerical results for the strip geometry 
are in agreement with the experiment of Heida {\it et
al.}~\cite{bib:Heida}. For wider junctions, we predict a crossover from a
$\Phi_{0}$- to a $2\Phi_{0}$-periodicity at high fields.

We thank D. Agterberg and V. Geshkenbein for stimulating discussions throughout this work.

\begin{table}
 \caption{The periodicity of the critical current is controlled by the three
 parameters $w/d$, $w/r$, and $r/a_{\scriptscriptstyle{0}}$. The table has to
 be read as a flow chart, starting at the top row and selecting the proper
 condition proceeding down the rows. Note that always $w/d<w/r$.}
 $$
 \offinterlineskip\tabskip=0pt \vbox{ \halign to 0.955\hsize {\strut \vrule
 width0.8pt\quad# & \hfil # \hfil \quad & \vrule# & \quad # \quad & \vrule# &
 \quad # \quad & \vrule# & \quad # \quad & \vrule# & \quad # \quad & \vrule
 width0.8pt# \cr \noalign{\hrule}\noalign{\hrule} & ratio && \multispan 7
 \hfill value \hfill & \cr \noalign{\hrule} & $w/d$ && $>1$ && \multispan 3
 \hfill $<1$ \hfill && $\rightarrow 0$ & \cr & $w/r$ && \multispan 3 \hfill
 $>1$ \hfill && $<1$ && &\cr & $r/a_{\scriptscriptstyle{0}}$ && $<1$ && $>1$
 && && &\cr \noalign{\hrule} & period && $\Phi_{0}$ && \multispan 3 \hfill
 $2\Phi_{0}$ \hfill && none & \cr \noalign{\hrule}\noalign{\hrule} }}
 $$
 \label{t:ptable}
\end{table}

\end{multicols}


\begin{thebibliography}{99}
\bibitem{bib:rb}In the opposite case $r>w>a_{\scriptscriptstyle{0}}$, we are
in the limit $w/d\rightarrow 0$, where the order-zero critical current is
given by the result conjectured in Ref. \cite{bib:Heida}.
\end{thebibliography}
\end{document}